\documentclass[twocolumn,showkeys,aps,prb,showpacs]{revtex4-1}
\usepackage{graphicx}
\usepackage[CJKbookmarks,dvipdfm,colorlinks,linkcolor=blue,citecolor=blue]{hyperref}

\begin{document}

\title{Tuning  transport coefficients of monolayer $\mathrm{MoSi_2N_4}$ with biaxial  strain}
\author{Xiao-Shu Guo$^{1}$ and San-Dong Guo$^{2,3}$}
\affiliation{$^1$Xi'an University of Posts and Telecommunications, Xi'an 710121, China}
\affiliation{$^2$School of Electronic Engineering, Xi'an University of Posts and Telecommunications, Xi'an 710121, China}
\affiliation{$^3$Key Laboratary of Advanced Semiconductor Devices and Materials, Xi'an University of Posts and Telecommunications, Xi'an 710121, China }
\begin{abstract}
Experimentally synthesized $\mathrm{MoSi_2N_4}$ (\textcolor[rgb]{0.00,0.00,1.00}{Science 369, 670-674 (2020)}) is a piezoelectric semiconductor.
Here, we systematically study the large biaxial (isotropic)  strain effects (0.90 to 1.10) on  electronic structures and transport coefficients of monolayer $\mathrm{MoSi_2N_4}$ by density functional theory (DFT). With $a/a_0$ from 0.90 to 1.10, the energy band gap firstly increases, and then decreases, which is due to transformation of  conduction band minimum (CBM). Calculated results show that the $\mathrm{MoSi_2N_4}$ monolayer is mechanically stable in considered strain range.
It is found that the spin-orbital coupling (SOC) effects on Seebeck coefficient depend on the strain. In unstrained $\mathrm{MoSi_2N_4}$,
 the SOC has neglected influence on Seebeck coefficient. However, the SOC can produce important influence on Seebeck coefficient, when the strain is applied, for example 0.96 strain.
The compressive strain can change relative position and numbers of conduction band extrema (CBE), and then the strength of conduction bands convergence  can be enhanced, to the benefit of n-type $ZT_e$.  Only about 0.96 strain can effectively improve  n-type $ZT_e$.
 Our works imply that strain can effectively
tune the  electronic structures and transport coefficients of monolayer $\mathrm{MoSi_2N_4}$, and can  motivate farther experimental exploration.

\end{abstract}
\keywords{$\mathrm{MoSi_2N_4}$, Electronic transport,  2D materials}

\pacs{71.20.-b, 72.15.Jf ~~~~~~~~~~~~~~~~~~~~~~~~~~~~~~~~~~~Email:sandongyuwang@163.com}

\maketitle

\section{Introduction}
The successful exfoliation of graphene\cite{q6}  induces increasing attention on  two-dimensional (2D)  materials.
Many of them have semiconducting behaviour, which has various  potential application in electronics, optoelectronics and piezoelectronics\cite{m4-1,m5-1,xzq,m2-1}. Their  electronic structures,   heat transport and piezoelectric properties  have been widely investigated\cite{m1,m2,m3,m4,m5,m6,m7,m8,m9,m10,m11}. It has been proved that the strain can effectively  tune electronic structures, transport and  piezoelectric properties of 2D materials\cite{m12,m10,m11,m13,m14,m15,m16,m17,m18}, which shows great potential for better use in the nanoelectronic, thermoelectric and piezoelectric applications. For example, both compressive and tensile strain can induce the semiconductor to metal transition in monolayer  $\mathrm{MoS_2}$\cite{m12}.   In many transition metal dichalchogenides (TMD) monolayers, the power factor can be enhanced by strain  due to bands converge\cite{m10,m11,m13}. With  increased tensile strain, the lattice thermal conductivity shows monotonous decrease,   up-and-down  and jump behavior with similar penta-structures\cite{m14}. Strain can also improve the piezoelectric  strain  coefficient by tuning the elastic and piezoelectric stress  coefficients\cite{m15,m16,m17,m18}.
\begin{figure}
  \includegraphics[width=8.0cm]{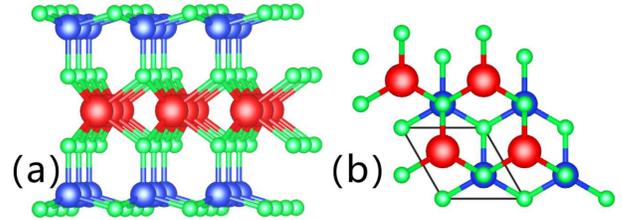}
  \caption{(Color online) The crystal structure of monolayer  $\mathrm{MoSi_2N_4}$ ((a) side view and (b) top view). The primitive cell is
   are marked by black  line, and  the large red balls represent Mo atoms, and the middle blue balls for Si atoms, and the  small green balls for N atoms. }\label{t0}
\end{figure}

Recently, the layered
2D $\mathrm{MoSi_2N_4}$ and $\mathrm{WSi_2N_4}$ have been  experimentally achieved by chemical vapor deposition (CVD)\cite{m19}. The septuple-atomic-layer $\mathrm{MA_2Z_4}$ monolayers with twelve different structures are constructed by intercalating $\mathrm{MoS_2}$-type  $\mathrm{MZ_2}$ monolayer into InSe-type  $\mathrm{A_2Z_2}$ monolayer\cite{m20}.  The 66 thermodynamically and
dynamically stable $\mathrm{MA_2Z_4}$  are predicted by the first principle calculations. They can be common semiconductor,  half-metal ferromagnetism or spin-gapless  semiconductor (SGS),  Ising superconductor and  topological insulator, which depends on the number of valence
electrons\cite{m20}. We predict intrinsic piezoelectricity in monolayer  $\mathrm{MA_2Z_4}$\cite{m21}, which means that  $\mathrm{MA_2Z_4}$  family may have potential application in piezoelectric field.   Structure  effect on intrinsic piezoelectricity in monolayer  $\mathrm{MSi_2N_4}$ (M=Mo and W) has also been reported by the first principle calculations\cite{m21-1}.
By applied strain, the $\mathrm{VSi_2P_4}$ monolayer undergoes  ferromagnetic metal (FMM) to SGS to ferromagnetic semiconductor (FMS) to SGS to ferromagnetic half-metal (FMHM) with increasing strain\cite{m22}.
Some materials of $\mathrm{MA_2Z_4}$  lack inversion symmetry with a strong SOC effect, which are expected to exhibit rich spin-valley physics\cite{m20}. The valley-dependent properties of monolayer $\mathrm{MoSi_2N_4}$, $\mathrm{WSi_2N_4}$ and $\mathrm{MoSi_2As_4}$ have been predicted by the first-principles calculations\cite{m20,m23,m24}. Recently, Janus 2D monolayer  in the new septuple-atomic-layer 2D $\mathrm{MA_2Z_4}$ family has been achieved\cite{m25}, which shows Rashba spin splitting and out-of-plane  piezoelectric polarizations.
\begin{figure*}
  \includegraphics[width=15cm]{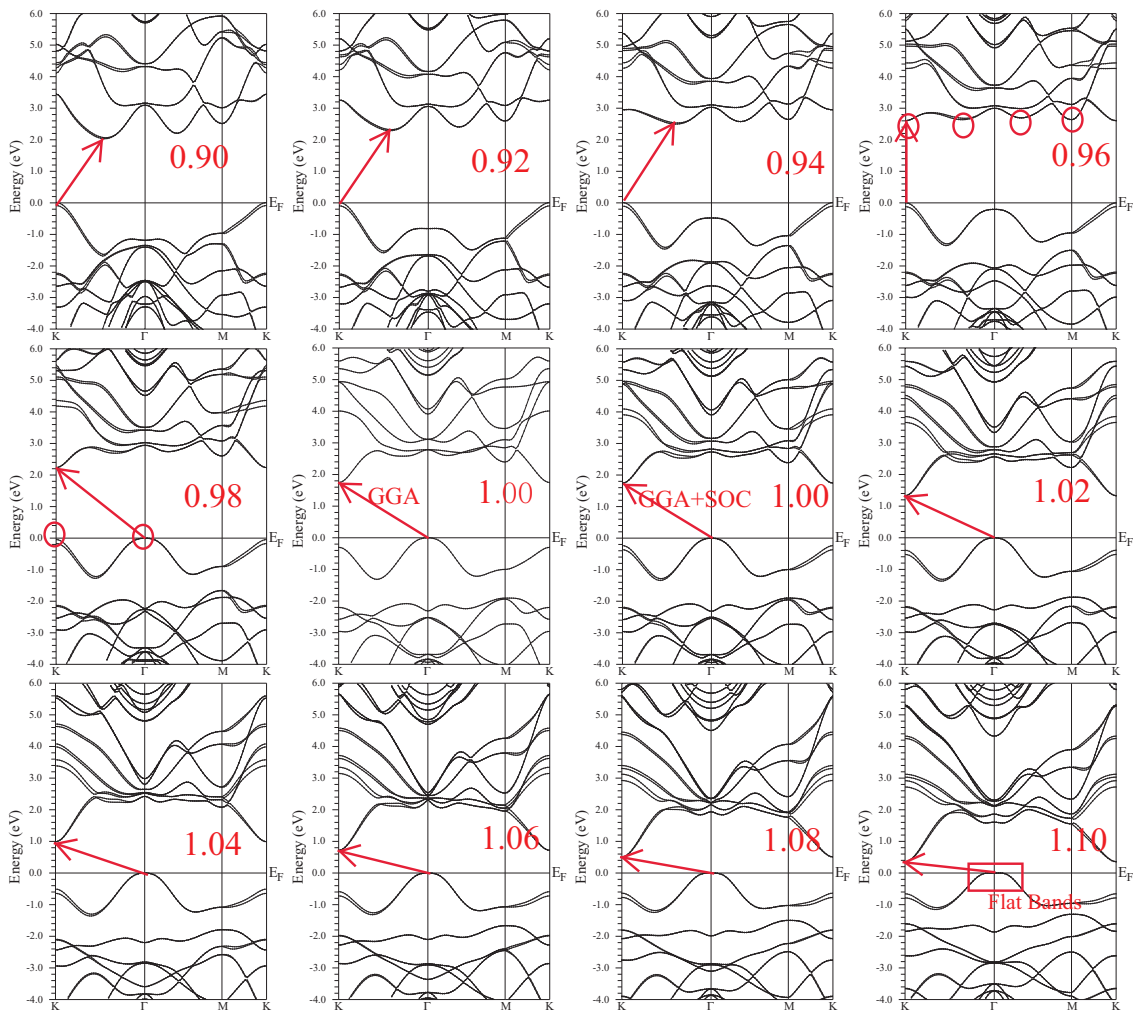}
  \caption{(Color online) The energy band structures  of   monolayer $\mathrm{MoSi_2N_4}$  using GGA+SOC  with the application of  biaxial strain (-10\% to 10\%), and the unstrained energy band using GGA. The VBM and CBM  are marked by arrows. At 0.96 (0.98) strain, four CBE (two VBE) are marked by ellipse.}\label{nd}
\end{figure*}
\begin{figure}
  \includegraphics[width=8cm]{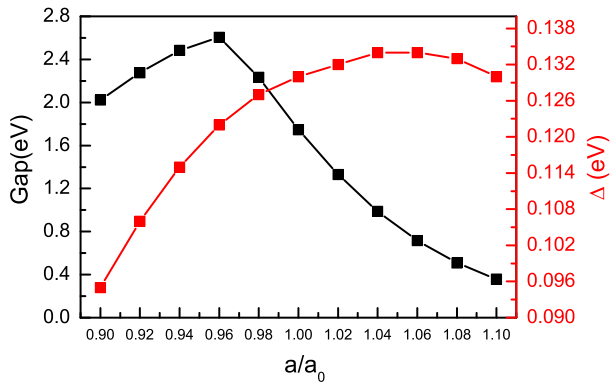}
  \caption{(Color online) For  $\mathrm{MoSi_2N_4}$ monolayer, the energy band gap  and  spin-orbit splitting value $\Delta$ at K point using GGA+SOC as a function of strain.}\label{nd1}
\end{figure}

In  nanoscale devices,  the  residual strain usually exists in real applications\cite{l111}. In our previous work, the small strain effects (0.96 to 1.04) on piezoelectric coefficients of monolayer $\mathrm{MoSi_2N_4}$  have been investigated\cite{m21}.
 In this work,  the large (0.90 to 1.10) biaxial  strain-tuned electronic structures and transport  coefficients of monolayer $\mathrm{MoSi_2N_4}$  are studied by the first principle calculations.  With $a/a_0$ from 0.90 to 1.10,  the energy band gap of monolayer $\mathrm{MoSi_2N_4}$ firstly increases, and then decreases. In n-type doping,  the  Seebeck coefficient S can be effectively enhanced  by applying  compressive strain, and then the $ZT_e$ can be improved. The tensile  strain can induce flat valence bands around the $\Gamma$ point near the Fermi level, producing large p-type S.
 Therefore, our works give an experimental proposal  to improve  transport coefficients of monolayer $\mathrm{MoSi_2N_4}$.

The rest of the paper is organized as follows. In the next
section, we shall give our computational details and methods  about transport  coefficients.
 In the third and fourth  sections, we will present main results of  monolayer $\mathrm{MoSi_2N_4}$ about strain-tuned electronic structures and transport  coefficients.  Finally, we shall give our  conclusions in the sixth section.

\section{Computational detail}
To avoid interactions
between two neighboring images, a vacuum spacing of
 more than 32 $\mathrm{{\AA}}$ along the z direction is added to construct monolayer  $\mathrm{MoSi_2N_4}$.
 The elastic stiffness tensor  $C_{ij}$   are calculated by using strain-stress relationship (SSR), which are performed  by using the  VASP code\cite{pv1,pv2,pv3}  within the framework of DFT\cite{1}. A kinetic cutoff energy of 500 eV is adopted, and  we use the popular generalized gradient
approximation  of Perdew, Burke and  Ernzerh of  (GGA-PBE)\cite{pbe} as the exchange-correlation potential to calculate elastic and electronic properties.
The total energy  convergence criterion is set
to $10^{-8}$ eV, and  the Hellmann-Feynman forces  on each atom are less than 0.0001 $\mathrm{eV.{\AA}^{-1}}$.
The Brillouin zone (BZ) sampling
is done using a Monkhorst-Pack mesh of 15$\times$15$\times$1  for elastic constants $C_{ij}$.
The 2D elastic coefficients $C^{2D}_{ij}$ have been renormalized by the the length of unit cell along z direction ($Lz$):  $C^{2D}_{ij}$=$Lz$$C^{3D}_{ij}$ .

The electronic transport coefficients of $\mathrm{MoSi_2N_4}$ monolayer  are calculated through solving Boltzmann
transport equations within the constant scattering time approximation (CSTA), which is performed by BoltzTrap\cite{b} code. To include  the SOC,  a full-potential linearized augmented-plane-waves method
  is used to calculate the energy bands  of   $\mathrm{MoSi_2N_4}$ monolayer, as implemented in
the WIEN2k  package\cite{2}. To attain  accurate transport coefficients, a 35 $\times$35$\times$ 1 k-point meshes is used in the first BZ for the energy band calculation, make harmonic expansion up to $\mathrm{l_{max} =10}$ in each of the atomic spheres, and set $\mathrm{R_{mt}*k_{max} = 8}$.

\section{Electronic structures}
The $\mathrm{MoSi_2N_4}$ monolayer can be  considered as the insertion of the 2H  $\mathrm{MoS_2}$-type  $\mathrm{MoN_2}$ monolayer into the  $\alpha$-InSe-type  $\mathrm{Si_2N_2}$, and the side and top views of the structure of the
$\mathrm{MoSi_2N_4}$ monolayer are plotted  in \autoref{t0}.
 The structure breaks the inversion symmetry, but preserves a horizontal mirror
corresponding to the plane of the Mo layer.  This leads to that $\mathrm{MoSi_2N_4}$ monolayer only has in-plane piezoelectric response, and has not out-of-plane  piezoelectric polarizations.
Using optimized lattice constants\cite{m21}, the energy bands of $\mathrm{MoSi_2N_4}$ monolayer using GGA and GGA+SOC are shown in \autoref{nd}, and exhibit both the indirect band gaps with valence band maximum (VBM) at $\Gamma$ point  and CBM at K point.
 Due to lacking inversion symmetry and containing  the heavy element Mo, there exists a  SOC induced spin splitting of about 0.13 eV  near the Fermi level in the valence bands  at K point.
This may provide a platform for spin-valley physics\cite{m20,m23,m24}, but the VBM is not at K point, which can be tuned by strain. According to orbital projected band structure, it is found  that the states near the Fermi level are
dominated by the Mo-$d$ orbitals. More specifically, the
states around both CBM and VBM are dominated by the Mo
$d_{z^2}$ orbital.
\begin{figure}
  \includegraphics[width=8cm]{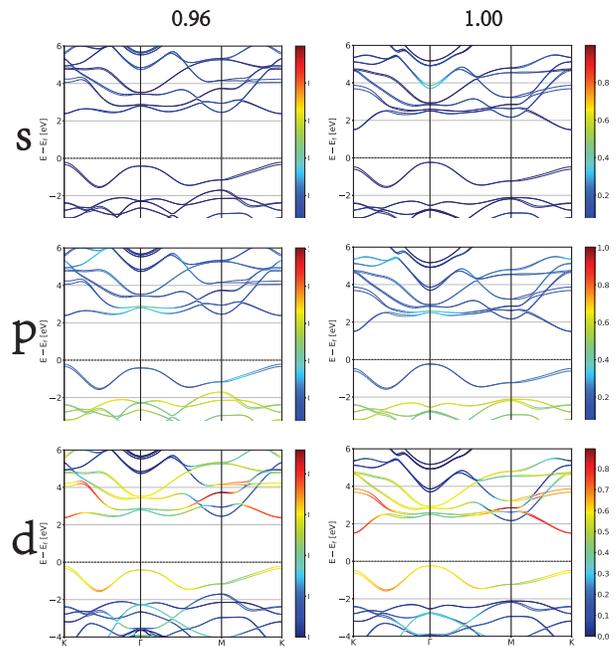}
  \caption{(Color online) For  $\mathrm{MoSi_2N_4}$ monolayer,  the orbital projected band structure at 0.96 strained and unstrained conditions.}\label{po}
\end{figure}

 It is proved that the electronic structures, topological properties, transport and   piezoelectric  properties of 2D materials can be effectively tuned by strain\cite{m12,m10,m11,m13,m14,m15,m16,m17,m18,t9}. The biaxial strain can be simulated by $a/a_0$ or $(a-a_0)/a_0$, where $a$ and $a_0$ are the strained and  unstrained lattice constant, respectively.  The $a/a_0$$<$1 or $(a-a_0)/a_0$$<$0 means  compressive strain, while  $a/a_0$$>$1 or $(a-a_0)/a_0$$>$0 implies tensile strain.   With $a/a_0$ from 0.90 to 1.10, the energy band structures  are plotted in \autoref{nd}, and
 the energy band gap  and  spin-orbit splitting value $\Delta$ at K point   are shown in \autoref{nd1}.

It is found that the energy band gap  firstly increases (0.90 to 0.96), and then decreases  (0.96 to 1.10), which is due to transformation of  CBM. Similar phenomenon can be observed in
 many TMD and Janus TMD monolayers\cite{m11,m11-1}. With strain from compressive one to tensile one, the $\Delta$ has a rapid increase, and then  a slight decrease. With increasing compressive strain (1.00 to 0.90), the position of CBM (VBM) changes from K ($\Gamma$) point to one point along the K-$\Gamma$ direction (K point),  when the compressive strain reaches about 0.94 (0.96). The compressive strain can also tune the numbers and relative positions of valence band extrema (VBE) or CBE. For example, at 0.96, the four CBE can be observed, and  they energies  are very close, which has very important effects on transport properties. To explore orbital contribution to the conduction bands in the case of 0.96 strain,  we project the
states to atomic orbitals at 0.96 strained and unstrained conditions, which  are shown in \autoref{po}. At 0.96 strain, the composition of the low-energy states has little change with respect to unstrained one.
 At 0.98, the energy of two VBE are  nearly the same. The compressive strain can  make  K point with spin splitting  become VBM, which is very useful  to allow spin manipulation for spin-valley physics. For example, at 0.94 strain, the VBM at K point  is 0.49 eV higher
than that at $\Gamma$ point.
It is clearly seen that the increasing  tensile strain  can make valence band  around the $\Gamma$ point near the Fermi level more flat.
\begin{figure}
  \includegraphics[width=8cm]{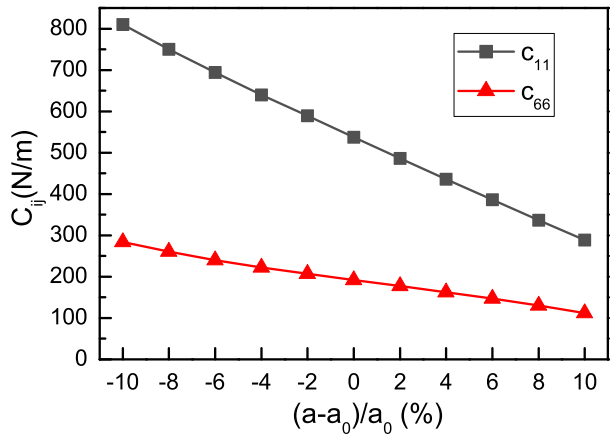}
  \caption{(Color online) For  $\mathrm{MoSi_2N_4}$ monolayer, the  elastic constants $C_{11}$ and $C_{66}$ vs  $a/a_0$ from 0.90 to 1.10.}\label{nd2}
\end{figure}

Finally,  the elastic constants $C_{ij}$ are calculated as a function of $a/a_0$ to study the mechanical  stability of $\mathrm{MoSi_2N_4}$ monolayer with strain. For for 2D hexagonal crystals, the  Born  criteria of mechanical stability \cite{ela} ($C_{11}>0$ and  $C_{66}>0$) should be satisfied.
The calculated $C_{11}$ and  $C_{66}$ as a function of strain are plotted in \autoref{nd2}, and it is clearly seen that the $\mathrm{MoSi_2N_4}$ monolayer  in considered strain range is mechanically stable, which is very important for farther experimental exploration.
\begin{figure}
  \includegraphics[width=8cm]{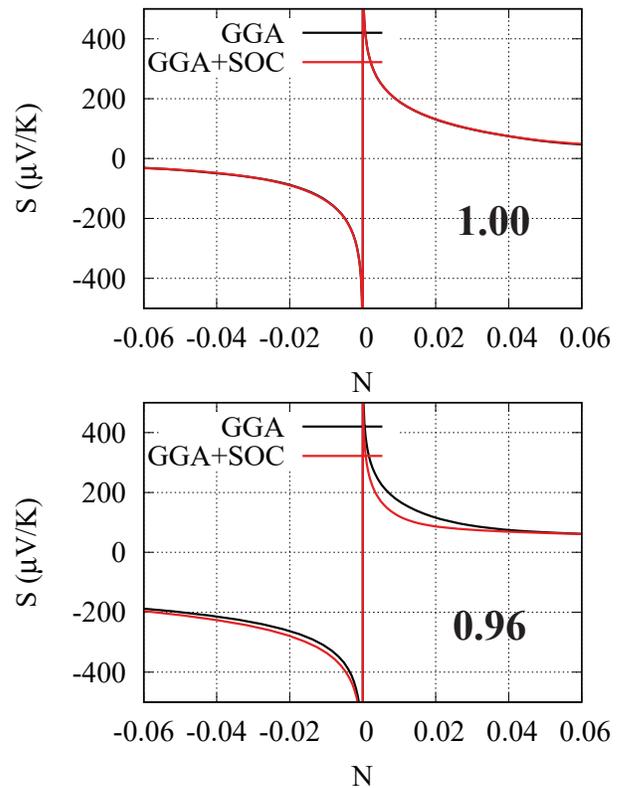}
  \caption{(Color online) For  $\mathrm{MoSi_2N_4}$ monolayer, the room-temperature Seebeck coefficient S using GGA and GGA+SOC at 1.00  and 0.96 strains  as a function of doping level N (The N means number of electrons or holes per unitcell).}\label{s1-1}
\end{figure}

\begin{figure*}
  \includegraphics[width=14cm]{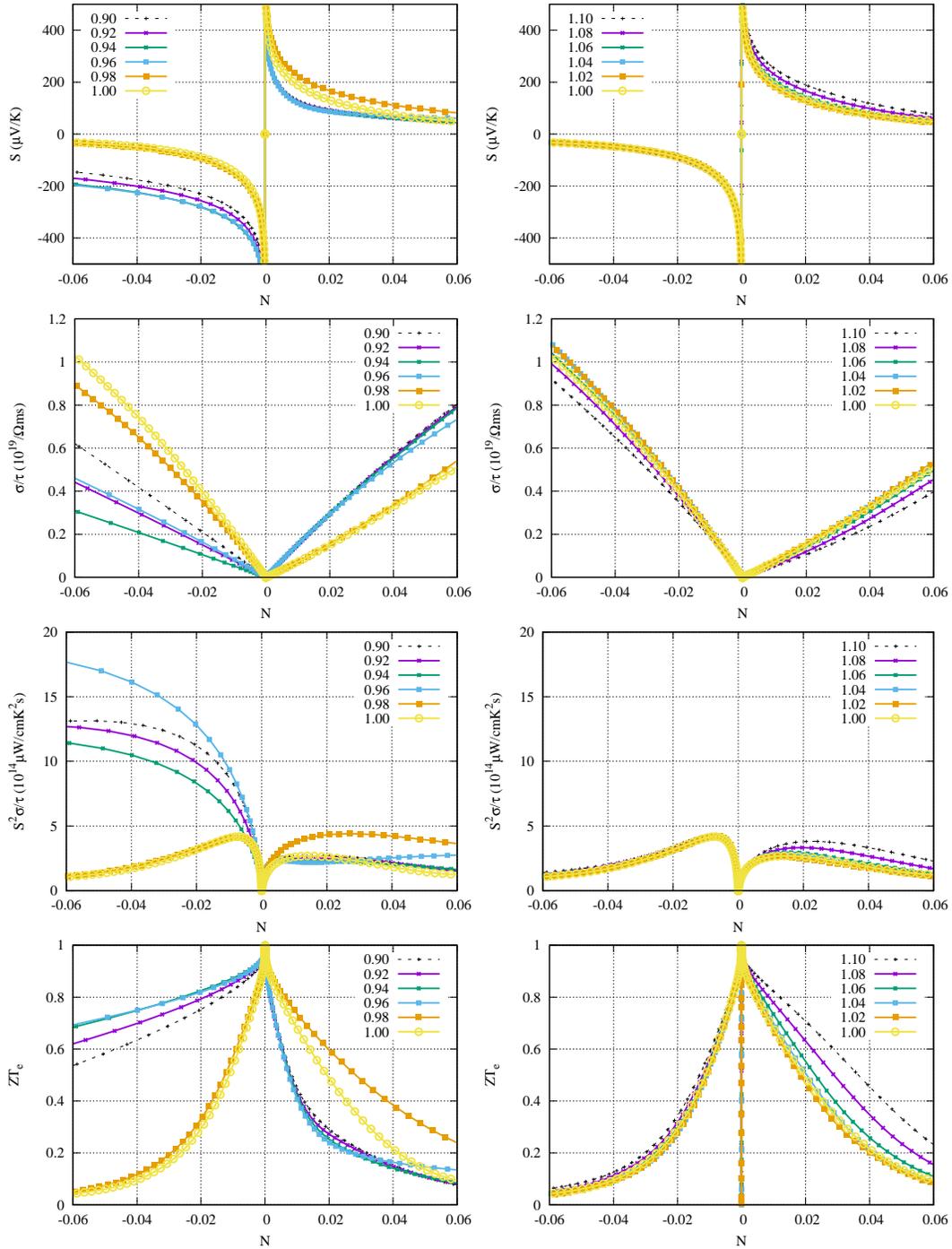}
  \caption{(Color online) For  $\mathrm{MoSi_2N_4}$ monolayer, the room-temperature transport coefficients   with the $a/a_0$  from 0.90 to 1.10 [(Left): compressive strain and (Right): tensile strain]: Seebeck coefficient S, electrical conductivity with respect to scattering time  $\mathrm{\sigma/\tau}$,   power factor with respect to scattering time $\mathrm{S^2\sigma/\tau}$ and $ZT_e$ (an upper limit of $ZT$) as a function of doping level (N) using GGA+SOC.}\label{s1}
\end{figure*}

\section{Electronic transport property}
Proposed by Hicks and Dresselhaus in 1993\cite{q2,q3},  the  potential thermoelectric materials can be achieved in  the low-dimensional
systems or nanostructures. The dimensionless  figure of merit, $ZT=S^2\sigma T/(\kappa_e+\kappa_L)$, can be used to measure
the efficiency of thermoelectric conversion of a thermoelectric material,  where  S, $\sigma$, T, $\kappa_e$ and $\kappa_L$ are the Seebeck coefficient, electrical conductivity, working temperature,  electronic and lattice thermal conductivities, respectively.
It is noted that, for the 2D material, the calculated $\sigma$, $\kappa_e$ and $\kappa_L$ depend on $Lz$ (here,  $Lz$=40 $\mathrm{{\AA}}$), and the S and $ZT$ is independent of $Lz$.
 For 2D materials, we use electrons or holes per unit cell instead of doping concentration, which is described by N,  and the N $<$($>$) 0 mean n- (p-) type doping.
It is proved that the SOC has important effects on transport coefficients of TMD and Janus TMD monolayers\cite{m11,m13,m11-1}. However, the SOC has neglectful influences on transport properties of unstrained $\mathrm{MoSi_2N_4}$ monolayer, which can be observed from typical Seebeck coefficient S in \autoref{s1-1}.
This is because the energy bands near the Fermi level between GGA and GGA+SOC is nearly the same. However, the SOC has an important effect on p-type transport coefficients with the condition of compressive strain. For example at 0.96 strain,  a  detrimental effect on  Seebeck coefficient S can be observed, when including SOC (See \autoref{s1-1}).
This  is because  the SOC can remove the
band degeneracy  near the VBM. So, the SOC is included to investigate the biaxial  strain effects on transport coefficients of $\mathrm{MoSi_2N_4}$ monolayer.

Using GGA+SOC, the room temperature  S,  $\mathrm{\sigma/\tau}$ and $\mathrm{S^2\sigma/\tau}$ of  $\mathrm{MoSi_2N_4}$  monolayer  under different strain (0.90 to 1.10)   are shown  in \autoref{s1}. It is clearly seen that the compressive strain has important effects on S, especially for n-type doping. However, the tensile strain produces small influences on S, especially for n-type S. These can be explained by strain-induced energy bands.
When the strain is less than or equal to about 0.98, the n-type S (absolute value) can be observably improved, which is due to  compressive strain-driven  accidental conduction band degeneracies, namely bands convergence. With expanding compressive strain, in the low doping,  the p-type S firstly increases, and has almost no change. This is because the valence bands convergence  can be observed at about 0.98, and then is removed (At 0.98, the energy of two VBE are  nearly the same, and only one VBE near the Fermi level can be observed with compressive strain from 0.96 to 0.90.). Foe considered tensile strain, the conduction bands near the Fermi level have little change, which leads to almost unchanged n-type S. When the strain changes from 1.00 to 1.10, the p-type S increases, which is due to tensile strain-induced  more flat valence bands around $\Gamma$ point near the Fermi level.
This can be understood by $S=\frac{8\pi^2K_B^2}{3eh^2}m^*T(\frac{\pi}{3n})^{2/3}$, in which   $m^*$, T and  $n$ is  the effective mass of
the carrier, temperature and  carrier concentration, respectively.  The flat bands can  produce very large effective mass of the carrier, which will lead to improved S.  It is found that the   strain  has nearly the opposite effects on $\mathrm{\sigma/\tau}$ with respect to S.
It is found that the compressive strain can dramatically improve  $\mathrm{S^2\sigma/\tau}$  due to the strain-enhanced S.

An upper limit of $ZT$ can be measured by   $ZT_e=S^2\sigma T/\kappa_e$, neglecting  the  $\kappa_L$.
The room temperature $ZT_e$ of $\mathrm{MoSi_2N_4}$ monolayer under different strain as a function of  doping level  are also  shown in \autoref{s1}.
Calculated results show  that the dependence of  $ZT_e$  is very similar to one of S (absolute value), which can be explained by
the Wiedemann-Franz law: $\kappa_e=L\sigma T$ ($L$ is the Lorenz number).  And then the $ZT_e$ can be reformulated by $ZT_e=S^2/L$.
Thus, the strain-induced  bands convergence  improves  S, which  is  beneficial to better $ZT_e$.

\section{Conclusion}
In summary, we investigate the biaxial  strain (0.90 to 1.10) effects on electronic structures and  transport  coefficients of monolayer $\mathrm{MoSi_2N_4}$  by the  reliable first-principles calculations.  With the strain from 0.90 to 1.10, the energy band gap of $\mathrm{MoSi_2N_4}$ monolayer shows a nonmonotonic behavior.
It is found that the SOC  has little effects on  transport coefficients of unstrained $\mathrm{MoSi_2N_4}$ in considered doping range due to  the hardly changed dispersion of bands near the Fermi level. However,  the SOC has very important influences on  transport properties of strained $\mathrm{MoSi_2N_4}$, for example 0.96 strain, which is due to the position change of VBM.
Calculated results show that compressive strain can tune the numbers and  relative positions of CBE, which can lead to enhanced n-type S, and then better n-type $ZT_e$. Our works may provide an idea to optimize the electronic structures and transport  properties of monolayer $\mathrm{MoSi_2N_4}$.

\section{Data availability}
The data that support the findings of this study are available from the corresponding author upon reasonable request.

\begin{acknowledgments}
This work is supported by the Natural Science Foundation of Shaanxi Provincial Department of Education (19JK0809). We are grateful to the Advanced Analysis and Computation Center of China University of Mining and Technology (CUMT) for the award of CPU hours and WIEN2k/VASP software to accomplish this work.
\end{acknowledgments}

\end{document}